\documentclass[sigconf]{acmart}

\AtBeginDocument{%
  \providecommand\BibTeX{{%
    \normalfont B\kern-0.5em{\scshape i\kern-0.25em b}\kern-0.8em\TeX}}}

\setcopyright{acmcopyright}
\copyrightyear{2018}
\acmYear{2018}
\acmDOI{XXXXXXX.XXXXXXX}

\acmBooktitle{Woodstock '18: ACM Symposium on Neural Gaze Detection,
 June 03--05, 2018, Woodstock, NY} 
\acmPrice{15.00}
\acmISBN{978-1-4503-XXXX-X/18/06}

\usepackage{graphicx}
\usepackage[caption=false,font=footnotesize]{subfig}
\usepackage{textcomp}
\usepackage{comment}
\usepackage{microtype}
\usepackage{algorithmicx}
\usepackage[linesnumbered, ruled, boxed]{algorithm2e}
\usepackage{graphicx}
\usepackage{epstopdf}
\usepackage{listings}
\usepackage{makecell,multirow}
\usepackage{url}
\usepackage{xcolor}
\usepackage{float}
\usepackage{verbatim}
\usepackage{balance}
\usepackage{pifont}
\usepackage{enumitem}
\usepackage{adjustbox}
\usepackage[symbol]{footmisc}
\usepackage{afterpage}
\usepackage[symbol]{footmisc}
\usepackage[thicklines]{cancel}
\usepackage{tcolorbox}
\usepackage{threeparttable}

\begin{document}

\title{COCO: Testing Code Generation Systems via Concretized Instructions}

\author{Ming Yan}
\affiliation{%
 \institution{College of Intelligence and Computing, Tianjin University}
 \country{China}
}
% \email{yanming@tju.edu.cn}

\author{Junjie Chen}
\affiliation{%
 \institution{College of Intelligence and Computing, Tianjin University}
 \country{China}
}
% \email{junjiechen@tju.edu.cn}

\author{Jie M. Zhang}
\affiliation{%
 \institution{King’s College London}
 \country{London, United Kingdom}
}
% \email{jie.zhang@kcl.ac.uk}

\author{Xuejie Cao}
\affiliation{%
 \institution{College of Intelligence and Computing, Tianjin University}
 \country{China}
}
% \email{caoxuejie@tju.edu.cn}

\author{Chen Yang}
\affiliation{%
 \institution{College of Intelligence and Computing, Tianjin University}
 \country{China}
}
% \email{yangchenyc@tju.edu.cn}

\author{Mark Harman}
\affiliation{%
 \institution{University College London}
 \country{London, United Kingdom}
}
% \email{mark.harman@ucl.ac.uk}

\newcommand{\tech}{COCO}
\newcommand{\baselinep}{PEGASUS}
\newcommand{\baselinet}{Translation-Pivoting}

\newcommand{\finding}[1]{ \begin{tcolorbox}#1\end{tcolorbox}}
\DeclareRobustCommand{\legendsquare}[1]{%
	\textcolor{#1}{\rule{4ex}{2ex}}%
}

% For spacing
\newcommand{\distance}{5pt}
\setlength{\textfloatsep}{\distance}%set distance between figure/tables on the top/bottom with text
\setlength{\floatsep}{\distance}%set distance between figures or tables
\setlength{\intextsep}{\distance}%set distance between figures/tables in text with text
\setlength{\dbltextfloatsep}{\distance} %distance between a figure/table spanning both columns and the text;
\setlength{\dblfloatsep}{\distance} %distance between two figures/tables spanning both columns.

\definecolor{top1}{gray}{0.55}
\definecolor{top2}{gray}{0.65}
\definecolor{top3}{gray}{0.75}
\definecolor{top4}{gray}{0.85}
\definecolor{top5}{gray}{0.9}

\settopmatter{printacmref=false} 
\renewcommand\footnotetextcopyrightpermission[1]{}

\begin{abstract}
Code generation systems have been extensively developed in recent years to generate source code based on natural language instructions. 
However, despite their advancements, these systems still face robustness issues where even slightly different instructions can result in significantly different code semantics.
Robustness is critical for code generation systems, as it can have significant impacts on software development, software quality, and trust in the generated code.
Although existing testing techniques for general text-to-text software can detect some robustness issues, they are limited in effectiveness due to ignoring the characteristics of code generation systems. 
In this work, we propose a novel technique \tech{} to test the robustness of code generation systems.
It exploits the usage scenario of code generation systems to make the original programming instruction more concrete by incorporating features known to be contained in the original code.
A robust system should maintain code semantics for the concretized instruction, and \tech{} detects robustness inconsistencies when it does not. 
We evaluated \tech{} on eight advanced code generation systems, including commercial tools such as Copilot and ChatGPT, using two widely-used datasets. 
Our results demonstrate the effectiveness of \tech{} in testing the robustness of code generation systems, outperforming two techniques adopted from general text-to-text software testing by 466.66\% and 104.02\%, respectively. 
Furthermore, concretized instructions generated by \tech{} can help reduce robustness inconsistencies by 18.35\% to 53.91\% through fine-tuning.
\end{abstract}


% \keywords{datasets, neural networks, gaze detection, text tagging}


\maketitle

\section{Introduction}
\label{sec:intro}

Automated code generation is to automatically generate source code conforming to a given programming instruction~\cite{Mastropaolo2023copilot}.
This area has garnered significant attention over the years, with considerable effort invested in its development~\cite{ling2016latent, rabinovich2017abstract, yin2017syntactic}.
In particular, Large Language Models (LLMs) have made substantial progress in this field recently, such as CodeX~\cite{chen2021codex}, Copilot~\cite{github_copilot}, and ChatGPT~\cite{openai_chatgpt}, which take code generation to new heights of practicality.
For example, according to a recent report~\cite{copilot_blog}, Copilot has written nearly 40\% of code in files where it is enabled, thereby allowing developers to focus on tackling more complex software development challenges.

It is indeed encouraging about the breakthrough brought by LLM-based code generation systems in terms of prediction performance, but their robustness, an equally critical property, is often overlooked.
As pointed out by the existing work~\cite{Mastropaolo2023copilot}, slight changes in natural language instructions could result in generating significantly different code in semantics. 
Such semantic differences between the code snippets from the original instruction and the changed instruction indicate poor capabilities in code generation,
because at least one of the two code snippets must be incorrect.  

In practice, it is common for different developers to describe the same programming instruction in various ways.
Additionally, even for the same developer, providing the same instruction for the same code generation task every time is not guaranteed. 
These can have a negative impact on the practical use of code generation systems, potentially slowing down the development process or even damaging the software quality. 
Hence, it is crucial to thoroughly test and enhance the robustness of code generation systems.

\begin{figure*}[ht!]
    \center
    \begin{tabular}{c}\hspace{-4.5mm}  \includegraphics[scale=0.63]{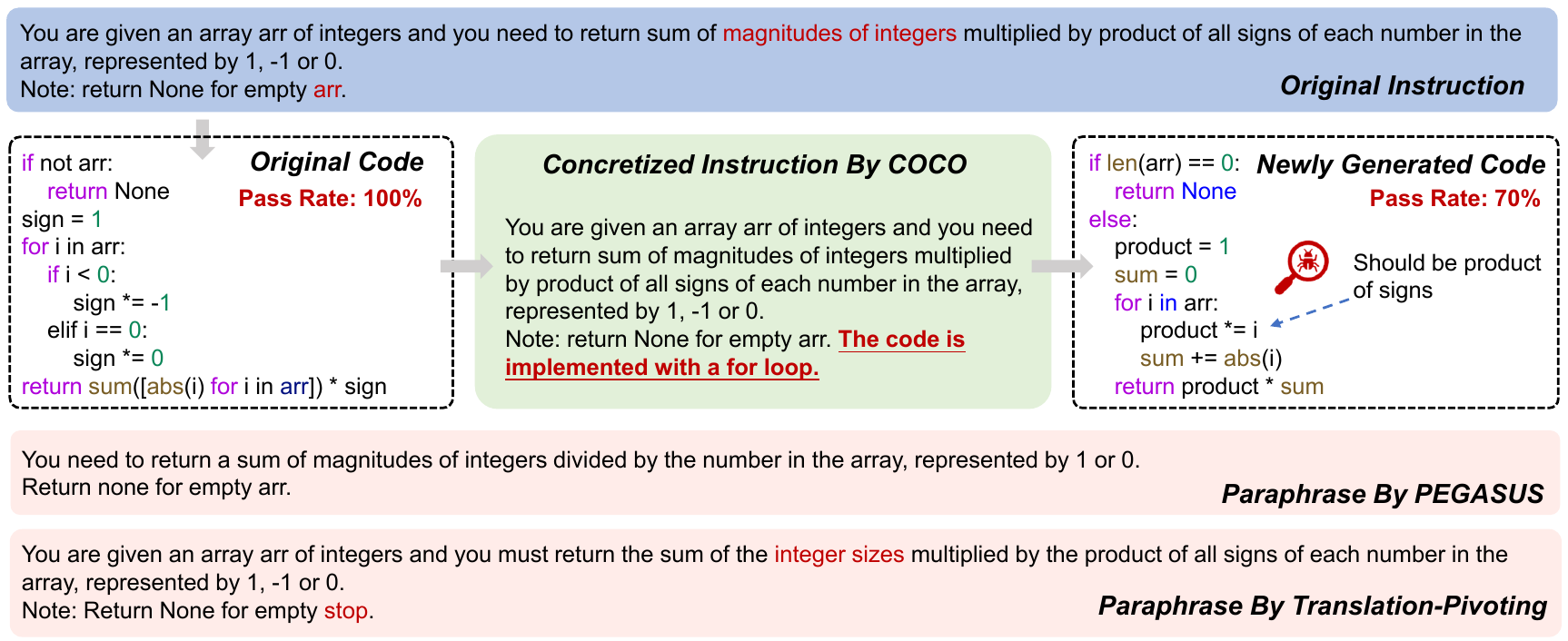}
    \end{tabular}
        \caption{A real example in our study from Copilot for the HumanEval dataset. }
    \label{fig:motivating}
\end{figure*}

There has been a recent study~\cite{Mastropaolo2023copilot} that conducted the first investigation on the robustness of code generation systems. 
They adopted the testing techniques for general text-to-text software (e.g., machine translators) to test the robustness of code generation systems, since both of them take natural language text as input.
Specifically, these general techniques paraphrase an instruction to semantic-preserving ones through deep learning or translation pivoting for testing the robustness of code generation systems~\cite{zhang2020pegasus,Zhou2018MT4MT}.
Indeed, they can detect some robustness issues in code generation systems.
Nevertheless, these techniques treat instructions as pure natural languages, thereby having limitations in handling programming language-related features in the instructions. 
For example, the instructions for code generation systems tend to contain programming-specific tokens (such as the variable identifier {\tt arr} as shown in Figure~\ref{fig:motivating}), which is often out of the capability of text-paraphrasing methods for generating semantic-preserving and natural changes.

In this work, we present a novel technique to test the robustness of code generation systems by exploiting the unique characteristics of code generation tasks.
We call this technique \textbf{\tech{}} (robustness testing of \textbf{CO}de generation systems via \textbf{CO}ncretizing instructions).
For each code generation task, \tech{} modifies the original instruction by adding requirements of code features that are already present in the originally generated code.
Specifically, 
\tech{} extracts code features from the code generated with the original instruction, and then transforms these features to natural language instructions (e.g., ``the code is implemented with loops'' or ``the code imports the {\tt NumPy} package''). 
These instructions are then incorporated into the original instruction.
The key insight is that such a change in the instruction is not expected to affect the generated code for robust code generation systems, given that the newly added requirement is already satisfied with the original instruction.

\tech{} consists of two major components:
(1) constructing concretized instructions by extracting and utilizing features from the code generated from an original instruction;
(2) checking the semantic consistency between the code generated by the concretized instruction and the code generated by the original instruction.
To test the robustness of code generation systems more thoroughly, we extract six levels of code features in \tech{} and design three types of consistency-checking mechanisms.
Unlike the above-mentioned general testing techniques that use paraphrasing, \tech{} provides more details based on the original instruction, avoiding limitations incurred by programming-specific tokens.
Furthermore, \tech{} treats the code generation system as a black box, and is not affected by any implementation details or model structure of the system.

We conducted an extensive study to evaluate \tech{} on eight advanced code generation systems with great diversity.
They are released and maintained by different organizations (e.g., Microsoft, OpenAI, and Salesforce), employ various neural network architectures (e.g., GPT-3.5 and CodeT5), and have different parameter scales (e.g., from 110M to 1.5B).
In particular, we studied two commercial systems (i.e., Copilot~\cite{github_copilot} and ChatGPT~\cite{openai_chatgpt} developed by OpenAI).
Then, we used the two most widely-used datasets, i.e., HumanEval~\cite{chen2021codex} and APPS~\cite{hendrycks2021apps}, as our code generation benchmarks.

Our results demonstrate the effectiveness of \tech{} in testing the robustness of code generation systems.
For example, \tech{} detected 466.66\% and 104.02\% more robustness inconsistencies than the two baseline techniques (\baselinep{} and \baselinet{}~\cite{Mastropaolo2023copilot}).
In addition, our results show that fine-tuning with the concretized instructions generated by \tech{} can significantly improve the robustness of code generation with a reduction of 18.35\% to 53.91\% robustness inconsistencies, without damaging the code-generation performance.

To sum up, this paper makes the following contributions:
\begin{itemize}

    \item We propose a novel technique (\tech{}) to test robustness of code generation systems. \tech{} concretizes the original programming instruction by incorporating features known to be contained in the code generated by the original instruction.

    \item To facilitate experimental replication and practical use, we have implemented and released \tech{}, which includes code feature extraction, concretized instruction construction, and code consistency checking. The project homepage can be found at~\cite{coco_homepage}. 

    \item We extensively evaluated \tech{} on eight advanced code generation systems (including two commercial systems: Copilot and ChatGPT) and two widely-used datasets. 
    The results showed \tech{} detected more robustness inconsistencies than the state-of-the-art baselines and can also help to enhance the robustness of code generation systems.
\end{itemize}

\section{Motivation}
Figure~\ref{fig:motivating} shows a simplified example from the HumanEval dataset. 
The original instruction describes the programming requirement. 
When applying \baselinep{} to it, the paraphrased instruction is non-semantic-preserving and omits important information such as ``multiplied by the product of all signs of each number''.
The potential reason lies in that it is a deep-learning-based text-paraphrasing technique and its effectiveness is limited by training data.
When applying \baselinet{} to it, it incorrectly translates ``magnitudes of integers'' to ``integer sizes'' and the identifier \texttt{arr} to ``stop''.
The reason is that the variable \texttt{arr} may be mistakenly recognized as an abbreviation for the word ``arrive'' when translated from English to French, which subsequently results in its translation to a similar word ``stop'' when translated back to English. This indicates that machine translators could struggle to handle programming-specific tokens well.

That is, these generated instructions cannot be directly used to test code generation systems without manual validation, as both of techniques cannot guarantee to generate equivalent instructions to the original one.

Recall the usage of code generation systems in practice, a developer describes the programming instruction (e.g., method definitions) to the system, and expects it to generate desired code.
Based on it, we propose \tech{} to mimic a more concrete programming instruction based on the original one.
In particular, the added details are known to be contained in the generated code corresponding to the original instruction.
On the one hand, it is helpful to design test oracles for detecting robust inconsistencies;
On the other hand, it ensures the validity of generated input instructions without paraphrasing the original instruction.
In this example, the state-of-the-art code generation system (i.e., Copilot) generates desired code for the original instruction, which passes all the test cases equipped with the example.
By analyzing the generated code, we can obtain one feature contained in it, i.e., a {\tt for} loop.
\tech{} transforms the code feature into a natural language sentence ``The code is implemented with a {\tt for} loop'', and then integrates it with the original one to form a new concretized instruction.
For a robust system, making the instruction more concrete in expectation should not produce semantically different code with the original one.
However, Copilot generates semantically different code, which just passes 70\% of test cases, and thus we found a robustness issue in it.

The example shows that the idea of concretizing instructions in \tech{} is effective to test robustness of code generation systems.
Moreover, it follows the natural usage of code generation systems and ensures validity of generated instructions.

\section{Approach}
\afterpage{\begin{figure*}[tp]
    \center
    \begin{tabular}{c}\hspace{-4.5mm}  \includegraphics[scale=0.73]{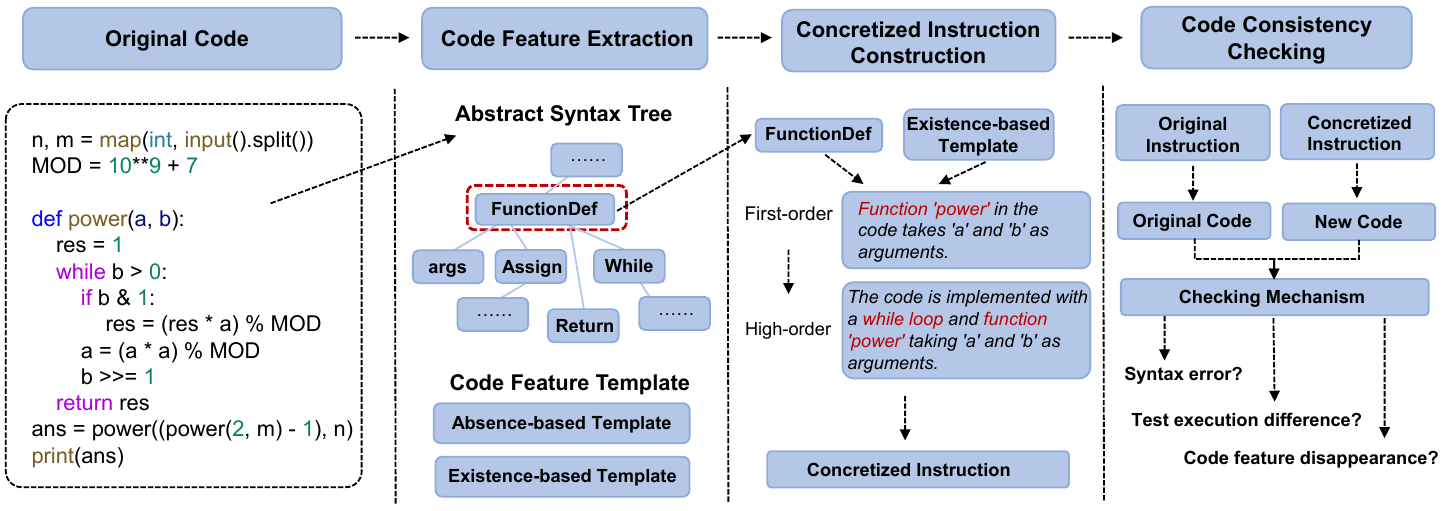}
    \end{tabular}
    \caption{Overview of \tech{}}
    \label{fig:overview}
\end{figure*}
}
The input instruction of code generation systems is to describe the programming requirement of the code to be generated.
That is, there is an implicit relationship between the input instruction and the generated code.
Based on this relationship, we design a novel technique for testing robustness of code generation systems, called \tech{}.
Instead of just relying on the original instruction to construct a semantic-preserving input, 
\tech{} utilizes the relationship between the input instruction and the generated code to guide the construction of a new input.
Specifically, \tech{} first extracts some features contained in the generated code corresponding to the original instruction, which actually captures some kind of implicit relationship between the input instruction and the generated code.
Then, it constructs a new input by transforming the features to a natural language instruction and integrating it with the original instruction.
That is, by making the extracted relationship more explicit, \tech{} makes the original programming instruction more concrete.
For a robust code generation system, the semantics of the generated code corresponding to the pair of the original instruction and the concretized instruction should be consistent.

In fact, \tech{} proposes a novel perspective to test robustness of code generation systems by closely exploiting the characteristics of this kind of systems.
Moreover, concretizing instructions conforms to the usage scenario of code generation systems by developers in practice, and thus can guarantee the naturalness of the constructed inputs by \tech{}.
Figure~\ref{fig:overview} shows the overview of \tech{}, which consists of three main steps: code feature extraction (to be presented in Section~\ref{code_feature}), concretized instruction construction (to be presented in Section~\ref{testing}), and code consistency checking (to be presented in Section~\ref{subsec:oracles}).

\subsection{Code Feature Extraction}
\label{code_feature}

In \tech{}, code features are extracted from the generated code corresponding to a given input instruction, which will be used to construct a concretized instruction for testing robustness of the code generation system.
In total, \tech{} extracts six levels of code features by systematically considering the programming-language syntax: dependency-level, class-level, method-level, statement-level, expression-level, and variable-level code features.
They involve different granularities of program elements, where the former three levels of code features are relatively coarse-grained and the latter three levels are relatively fine-grained.
The current \tech{} is implemented specific to Python code since Python has been demonstrated to be one of the most widely-used programming languages in practice~\cite{top_lang,github_top_lang}.
However, the idea of \tech{} is general and it can be easily generalized to other object-oriented programming languages (such as Java) since the six levels of code features are general to them.
In particular, \tech{} can also be generalized to other programming languages as long as designing corresponding code features.
In the following, we introduce each level of code features in details:
\begin{itemize}
    \item \textit{Dependency-level code features}: The generated code may import some libraries (such as {\tt NumPy} and {\tt Keras}) as dependencies for supporting the implementation of some functionalities.
    \tech{} extracts all the used libraries in the generated code, and a code feature at this level refers to whether the generated code depends on a certain library.

    \item \textit{Class-level code features}: The generated code may define some classes.
    \tech{} extracts the class information in the generated code, and a code feature at this level refers to whether the generated code defines a certain class.

    \item \textit{Method-level code features}: The generated code may define some methods, each of which may contain a return type and arguments.
    \tech{} extracts all this information for each method defined in the generated code.
    A code feature at this level refers to (1) whether the generated code defines a certain method, (2) whether a specific method has a certain return type, and (3) whether a specific method has a certain argument.

    \item \textit{Statement-level code features}:
    The generated code may use some types of statements, such as {\tt while} loops or {\tt switch} branches.
    \tech{} extracts all the statement types occurring in the generated code, and a code feature at this level refers to whether the generated code is implemented with a certain type of statements.

    \item \textit{Expression-level code features}:
    The generated code may use some types of expressions, such as logical expressions or arithmetic expressions.
    \tech{} extracts all the expression types occurring in the generated code, and a code feature at this level refers to whether the generated code is implemented with a certain type of expressions.

    \item \textit{Variable-level code features}:
    The generated code may define some variables.
    \tech{} extracts all the defined variables in the generated code, and a code feature at this level refers to whether the generated code defines a certain variable.

\end{itemize}

These code features are helpful to ensure the naturalness of generated instructions.
First, given the wide range of libraries available, developers may specify the used libraries in input instructions based on their programming habits or functionality requirements. 
Second, to make the generated code more coherent with the existing code in a project, developers may provide more details about the classes, methods, and variables to be generated in instructions.
Third, developers may also provide more concrete programming instructions (such as the statement and expression types to be used) to expect code generation systems performing better.

We take the code in Figure~\ref{fig:overview} as an example to illustrate the feature extraction process.
\tech{} first represents the code as an Abstract Syntax Tree (AST) and then visits AST nodes in the depth-first order to extract code features. 
During the visiting process, \tech{} defines a triple to represent each code feature. 
For example, for the \texttt{FunctionDef} node regarding \texttt{def power(a,b)}, \tech{} constructs a triple <Method, FunctionDef, power> to represent it.
This triple indicates that it is a method-level code feature with the definition of the method \texttt{power}.
The code feature extraction process terminates until the AST traversal is completed.

\subsection{Concretized Instruction Construction}\label{testing}

\newcommand{\mline}[1]{\uline{Line#1}}

A code generation system models the implicit relationship between an input instruction and the generated code.
Hence, the code features extracted from the generated code are actually implicitly indicated by the input instruction.
By making the indication appear in the input instruction more explicitly, the semantics of the generated code should keep consistent with the original one for a robust code generation system.
Based on this insight, \tech{} transforms the extracted code features to natural language instructions and then integrates them with the original input.
In this way, a concretized instruction can be constructed by incorporating more concrete programming requirements.

To facilitate transformation of code features to natural language text, we design natural language templates for different levels of code features in \tech{}.
In particular, we consider two types of templates to show the \textit{existence} of a code feature and the \textit{absence} of a code feature, respectively. 
For example, given the code feature from {\tt FunctionDef} in Figure~\ref{fig:overview}, \tech{} can use an existence-based template to transform the code feature into a natural language instruction:
``Function `power' in the code takes `a' and `b' as arguments''. 
For the original code in Figure~\ref{fig:overview}, it does not contain any class definitions, and thus we can use an absence-based template to transform the statement-level code feature into a natural language instruction: ``The code is implemented without classes''.

When constructing a concretized instruction, \tech{} has larger search space than the existing techniques~\cite{Mastropaolo2023copilot}.
For a given instruction, larger search space provides \tech{} more opportunities to generate instructions effective to trigger robustness inconsistencies, and thus improves the overall test effectiveness.
Specifically, for a given input, \tech{} has at most six levels of code features that can be applicable to construct concretized instructions since not all generated code contains all these code features.
When specifying one level of code features, there may exist several specific code features (e.g., there are both {\tt Assign} and {\tt While} in Figure~\ref{fig:overview} belonging to statement-level code features).
Based on template-based transformation, each specific code feature can be used to construct two concretized instructions by utilizing existence-based and absence-based templates respectively.
Furthermore, a concretized instruction can be constructed in a high-order way by assembling multiple code features from various levels. 
For example, \tech{} can generate a second-order concretized instruction by considering a method-level code feature from {\tt FunctionDef} and a statement-level code feature from {\tt While} together based on an existence-based template.

Due to the large search space, \tech{} cannot enumerate the entire space for concretized instruction construction and thus performs sampling to balance test effectiveness and efficiency.
By default, \tech{} conducts first-order concretization for constructing a concretized instruction.
For each level of code features, \tech{} randomly samples $m$ specific code features, each of which is used to construct a concretized instruction.
In the study, we will investigate the influence of $m$ on the effectiveness of \tech{}, and also compare the effectiveness of \tech{} based on first-order concretization and high-order concretization (denoted as $n$-order concretization).
Note that the key contribution of \tech{} is to construct concretized instructions by extracting code features from the generated code corresponding to a given instruction, rather than optimize the search process.
In the future, we can design more advanced sampling algorithms to improve the effectiveness of \tech{}.

\subsection{Code Consistency Checking}\label{subsec:oracles}

Based on a pair of the original instruction and the concretized instruction, \tech{} tests the code generation system and checks whether their generated code is semantically consistent.
However, checking semantic consistency automatically is quite challenging.
To more sufficiently capture robustness issues, \tech{} designs three types of checking mechanisms.

\begin{itemize}
    \item \textit{Syntax error}: If the generated code corresponding to the original instruction does not have syntax errors but that corresponding to the concretized one has, an inconsistency is detected.
    \item \textit{Test execution difference}: When using the same test suite to execute the pair of generated code, if there are test execution differences, an inconsistency is detected.
    In \tech{}, we used the number of passing test cases to measure test execution differences following the existing work~\cite{chen2021codex}.
    \item \textit{Code feature disappearance}: If an extracted code feature disappears in the code generated by the concretized instruction, an inconsistency is detected.

\end{itemize}

In addition, the existing study~\cite{Mastropaolo2023copilot} used CodeBLEU~\cite{Ren2020CodeBLEU} to measure textual similarity between generated code in robustness testing of code generation systems, which is more fuzzy as textual differences may not indicate semantic inconsistencies. 
Therefore, we did not integrate it into \tech{}, but 
discussed the results measured by it as complementary in Section~\ref{sec:discussion}.

\section{Experimental Design}
We evaluated \tech{} by answering the following research questions (RQs):

\noindent\textbf{RQ1: Can \tech{} detect robustness inconsistencies in code generation systems effectively?}

To answer RQ1, we applied and evaluated \tech{} and two baselines on all subjects. Since the baselines may produce invalid inputs, we manually analyzed and filtered out all false positives generated by the baselines. To evaluate the ability of \tech{} and the baselines to detect robustness inconsistencies, we further compared the number of inconsistencies detected by each technique and analyzed the overlap of the detected inconsistencies.

\noindent\textbf{RQ2: How do each level of code features and each type of templates contribute to the effectiveness of \tech{}?}

To answer RQ2, we evaluated the effectiveness of each level of code features and two types of templates in \tech{}. 
We compared the number of inconsistencies detected by each level of code features and each type of templates, respectively.

\noindent\textbf{RQ3: Can \tech{} help enhance the robustness of code generation systems?}

To answer RQ3, we adopted the widely-used fine-tuning strategy to enhance the robustness of code generation systems with the generated instructions by \tech{}.

\subsection{Code Generation Systems}
We used eight well-known code generation systems in our study to evaluate \tech{}.
These subjects are diverse:
(1) We consider the code generation systems from different organizations, including two commercial systems (i.e., Copilot~\cite{github_copilot} and ChatGPT~\cite{openai_chatgpt} developed by OpenAI) and six open-source systems (i.e., CodeGen~\cite{nijkamp2022codegen} and CodeRL~\cite{le2022coderl} hosted by Salesforce, InCoder~\cite{fried2022incoder} hosted by Facebook, PyCodeGPT~\cite{CERT} hosted by Microsoft, GPT-2 1.5B~\cite{radford2019language} and GPT-2 117M~\cite{radford2019language} hosted by OpenAI).
(2) We consider various mainstream neural network architectures, including GPT-Neo (i.e., PyCodeGPT), GPT-2 (i.e., GPT-2 1.5B, GPT-2 117M), GPT-3 (i.e., Copilot), GPT-3.5 (i.e., ChatGPT), CodeT5 (i.e., CodeRL), CodeGen, and InCoder.
(3) We consider different parameter scales of LLMs: small scales (e.g., 350M for CodeGen), medium scales (e.g., 1B for InCoder), and large scales for ChatGPT and Copilot. 
We did not know the specific parameter scales for Copilot and ChatGPT since the internal information of the two commercial systems is unavailable.

In particular, both Copilot and ChatGPT received extensive attention from academia and industry recently, which can be regarded as the state-of-the-art code generation systems.
Specifically, Copilot is especially designed for the task of code generation and is developed as a well-known programming auto-completion tool for IDEs.
ChatGPT is more general and code generation is one of the tasks that it can handle well.
The sampling strategies in code generation systems (such as the temperature sampling strategy used by ChatGPT) can incur non-determinism, which can affect the comparison between \tech{} and baselines.
Therefore, following the existing work~\cite{li2022cctest}, we disabled sampling strategies in code generation systems to eliminate the influence of non-determinism.

\begin{table}[t]
    \caption{\label{tab:models} Statistical information of our subjects}
    \begin{adjustbox}{max width=0.5\textwidth,center}
        \begin{threeparttable}
        \begin{tabular}{llccc}
        \toprule
        \textbf{ID.} & \textbf{Models} & \textbf{\# Parameters} & \textbf{Datasets} & \textbf{\# Size} \\
        \midrule
        1 & CodeGen-Multi & 350M & \multirow{5}{*}{HumanEval} & \multirow{5}{*}{164} \\
        2 & InCoder & 1B &  &  \\
        3 & PyCodeGPT & 110M &  &  \\
        4 & Copilot & --- &  &  \\
        5 & ChatGPT & --- &  &  \\ \midrule
        6 & CodeRL & 770M & \multirow{5}{*}{APPS} & \multirow{5}{*}{10,000} \\
        7 & GPT-2 & 1.5B &  &  \\
        8 & GPT-2 & 117M &  &  \\
        9 & Copilot & --- &  &  \\
        10 & ChatGPT & --- &  & \\ \bottomrule
        \end{tabular}
    \end{threeparttable}
    \end{adjustbox}
    \end{table}

\subsection{Code Generation Datasets} 
We used the two datasets, i.e., \textit{HumanEval}~\cite{chen2021codex} and \textit{APPS}~\cite{hendrycks2021apps}, as the test sets of code generation systems in our study.
That is, they provide original instructions for \tech{} and baselines.
Both of them have been widely used on the task of code generation~\cite{zhang2023planning,le2022coderl,CERT}.

HumanEval has been widely used to measure the functional correctness of generated code~\cite{li2022competition,chen2021codex,Xu2022systematic}.
It contains 164 hand-written programming problems.
Each problem includes a programming problem description and a set of test cases.
In total, there are more than 1,500 test cases. 
However, such a limited number of test cases may be insufficient for comprehensively evaluating the correctness of the generated code. Hence, Li et al.~\cite{liu2023code} augmented test cases for the HumanEval dataset and provided over 81 times more test cases on average for each problem. In this study, we used this augmented set with 122,487 test cases.
APPS contains 10,000 programming problems collected from different open-access coding websites (e.g., Codeforces and Kattis), where 5,000 problems are used as testing data.
Each problem also includes a programming problem description and a set of test cases.
For the 5,000 problems used as testing data, there are more than 106,000 test cases in total.
As presented in the existing work~\cite{chen2021codex}, the two datasets are complementary to some degree.
Table~\ref{tab:models} shows the combination of code generation systems and datasets used in our study similar to the existing work~\cite{Mastropaolo2023copilot,li2022cctest}.
Regarding the two commercial systems, we applied both datasets to them.
In total, we have 10 combinations for evaluation.

\subsection{Baseline Techniques}

In the literature, there is no testing techniques specific to code generation systems.
The existing work~\cite{Mastropaolo2023copilot} is the first to investigate the robustness of code generation systems by adopting the testing techniques for general text-to-text software.
They are \textbf{PEGASUS}~\cite{zhang2020pegasus} and \textbf{Translation-Pivoting} (TP)~\cite{Zhou2018MT4MT}.
The former is based on a pre-trained sequence-to-sequence model that can handle various natural language processing tasks, such as text paraphrasing.
Here, PEGASUS paraphrases the given original instruction to a new one, which is expected to be semantic-preserving with the original one.
The latter aims to construct a semantic-preserving input by translating the original text into the text of a pivot language and then translating it back to the text of the original language.

\subsection{Implementation and Configurations}
\label{subsec:implementations}
We implemented \tech{} based on Python 3.10, transformers 4.25.1, and openai 0.27.1 (the latest version when we began the study). 
\tech{} used the built-in Python package \texttt{ast}~\cite{python_ast} to construct and traverse the AST of the generated code.
The code features existing in the AST are transformed based on existence-based templates, while the remaining within the complete set of code features are transformed based on absence-based templates.
In \tech{}, we set $m$ to 1 by default, which carefully considers the test efficiency.
Moreover, we also investigated the effectiveness of \tech{} under different settings of $m$, i.e., $m=\{1,2,3,4,5\}$.
Similarly, \tech{} performs first-order concretization ($n=1$) by default for generating concretized instructions and investigated the effectiveness of \tech{} under different settings of $n$, i.e., $n=\{1,2,3,4,5\}$.
The influence of the two parameters will be presented in Section~\ref{subsec:parameters}.
Regarding baselines, we directly adopted their open-source implementations with the recommended configurations for comparison~\cite{robustness-copilot}.

Our experiments were conducted on an Intel Xeon CPU Gold-6342 machine with 512 GB RAM, Ubuntu 20.04.6 and four A800 GPUs. 
To reduce the influence of randomness, we repeated our experiment five times with different random seeds and confirmed the stability of \tech{}.
The details will be presented in Section~\ref{sec:threats}.

\subsection{Measurements}\label{subsec:metrics}
In the study, we used the number of detected inconsistencies to measure the effectiveness of \tech{} and the two baselines, which are based on our designed three checking mechanisms presented in Section~\ref{subsec:oracles}. Larger number of detected inconsistencies indicate better test effectiveness. 
Note that the checking mechanism of code feature disappearance is not applicable to both baselines since they do not consider code features.
Following the existing work~\cite{shen2022qaqa}, the inconsistencies, which are detected by several instructions generated from the same original instruction, are regarded as duplicate inconsistencies.
Hence, we de-duplicated them when measuring the effectiveness of each studied technique.

\section{Results and Analysis}
\label{sec:results}

\subsection{RQ1: Overall Effectiveness of \tech{}}\label{subsec:effectiveness}

\begin{table}[t]
    \caption{\label{tab:input_fps}Ratio of false positives produced by \baselinep{}, \baselinet{} and \tech{}.}
    \vspace{-1mm}
    \begin{adjustbox}{width=0.35\textwidth,center}
    \begin{threeparttable}

    \begin{tabular}{lrrr}
    \toprule
    \textbf{Dataset} & \textbf{PEGASUS} & \textbf{TP} & \textbf{\tech{}} \\
    \midrule
    HumanEval & 57.32\% & 11.59\% & 1.34\% \\
    APPS & 84\% & 13.30\% & 0.13\% \\ \midrule
    \end{tabular}

    \end{threeparttable}
    \end{adjustbox}
\end{table}
% \begin{table}[]
% \begin{tabular}{lllllllllll}
% \multirow{2}{*}{\textbf{Checking Mechanism}} & \multicolumn{5}{c}{\textbf{HumanEval}} & \multicolumn{5}{c}{\textbf{APPS}} \\
%  & \textbf{CodeGen} & \textbf{InCoder} & \textbf{PyCodeGPT} & \textbf{Copilot} & \textbf{ChatGPT} & \textbf{CodeRL} & \textbf{GPT2(1.5B)} & \textbf{GPT2(117M)} & \textbf{Copilot} & \textbf{ChatGPT} \\
% \textbf{Syntax Error} & 22/3/9 & 48/10/15 & 23/7/6 & 14/2/1 & 1/2/1 & 7/1/1 & 15/0/4 & 29/1/11 & 51/1/16 & 1/0/1 \\
% \textbf{Test Execution Difference} & 105/26/39 & 86/15/36 & 81/21/29 & 118/24/46 & 73/22/36 & 103/14/65 & 107/17/65 & 88/16/45 & 98/13/67 & 119/16/66 \\
% \textbf{Code Feature Disappearance} & 42/0/0 & 47/0/0 & 41/0/0 & 76/0/0 & 18/0/0 & 34/0/0 & 68/0/0 & 30/0/0 & 42/0/0 & 22/0/0 \\
% \textbf{Aggregation} & 114/29/48 & 103/25/51 & 93/28/35 & 122/26/47 & 73/24/37 & 109/15/66 & 128/17/69 & 113/17/56 & 118/14/83 & 123/16/67
% \end{tabular}
% \end{table}

\begin{table*}[t]
    \caption{\label{tab:rq1-effectiveness}Number of robustness inconsistencies detected by \tech{}, \baselinep{}, and \baselinet{}, respectively.}

    % \vspace{-1mm}
    \begin{adjustbox}{width=1.0\textwidth,center}
    \begin{threeparttable}
    \begin{tabular}{l|lllll|lllll}
    \toprule
    \multirow{2}{*}{\textbf{Checking Mechanism}} & \multicolumn{5}{c|}{\textbf{HumanEval}} & \multicolumn{5}{c}{\textbf{APPS}} \\
     & \multicolumn{1}{c}{\textbf{CodeGen}} & \multicolumn{1}{c}{\textbf{InCoder}} & \multicolumn{1}{c}{\textbf{PyCodeGPT}} & \multicolumn{1}{c}{\textbf{Copilot}} & \multicolumn{1}{c|}{\textbf{ChatGPT}} & \multicolumn{1}{c}{\textbf{CodeRL}} & \multicolumn{1}{c}{\textbf{GPT2 (1.5B)}} & \multicolumn{1}{c}{\textbf{GPT2 (117M)}} & \multicolumn{1}{c}{\textbf{Copilot}} & \multicolumn{1}{c}{\textbf{ChatGPT}} \\ \midrule
    \textbf{Syntax Error} & \colorbox{top3}{22}/3/9 & \colorbox{top3}{48}/10/15 & \colorbox{top3}{23}/7/6 & \colorbox{top3}{14}/2/1 & 1/\colorbox{top3}{2}/1 & \colorbox{top3}{7}/1/1 & \colorbox{top3}{15}/0/4 & \colorbox{top3}{29}/1/11 & \colorbox{top3}{51}/1/16 & \colorbox{top3}{1}/0/1 \\
    
    \textbf{Test Execution Difference} & \colorbox{top3}{105}/26/39 & \colorbox{top3}{86}/15/36 & \colorbox{top3}{81}/21/29 & \colorbox{top3}{118}/24/46 & \colorbox{top3}{73}/22/36 & \colorbox{top3}{103}/14/65 & \colorbox{top3}{107}/17/65 & \colorbox{top3}{88}/16/45 & \colorbox{top3}{98}/13/67 & \colorbox{top3}{119}/16/66 \\
    
    \textbf{Code Feature Disappearance} & \colorbox{top3}{42}/0/0 & \colorbox{top3}{47}/0/0 & \colorbox{top3}{41}/0/0 & \colorbox{top3}{76}/0/0 & \colorbox{top3}{18}/0/0 & \colorbox{top3}{34}/0/0 & \colorbox{top3}{68}/0/0 & \colorbox{top3}{30}/0/0 & \colorbox{top3}{42}/0/0 & \colorbox{top3}{22}/0/0 \\
    
    \textbf{Aggregation} & \colorbox{top3}{114}/29/48 & \colorbox{top3}{103}/25/51 & \colorbox{top3}{93}/28/35 & \colorbox{top3}{122}/26/47 & \colorbox{top3}{73}/24/37 & \colorbox{top3}{109}/15/66 & \colorbox{top3}{128}/17/69 & \colorbox{top3}{113}/17/56 & \colorbox{top3}{118}/14/83 & \colorbox{top3}{123}/16/67\\ \bottomrule
\end{tabular}
        \begin{tablenotes}
            \small
            \item[1] Each cell in the table represents the number of robustness inconsistencies found by \tech{}/\baselinep{}/\baselinet{} under different checking mechanisms.
            \item[2] \legendsquare{top3} indicates the result corresponding to the technique with the largest number of detected inconsistencies.
        \end{tablenotes}
\end{threeparttable}
\end{adjustbox}
\end{table*}

\subsubsection{Setup} 
To answer RQ1, we applied \tech{} and the two baselines (\baselinep{} and TP) to each subject and measured their effectiveness in terms of the number of detected inconsistencies.
The two baselines may produce invalid input instructions (i.e., instructions that have semantically different requirements from the original one), which will yield false positives in the detected inconsistencies.
The checking mechanism of code feature disappearance may also produce false positives in the detected inconsistencies by \tech{}, since semantically equivalent code without the code features indicated in the concretized instruction may be generated in actual.
Hence, before comparing the number of inconsistencies, 
we filtered out the false positives produced by each technique.
Specifically, we manually checked (1) whether the instructions generated by baselines introduce semantic changes (i.e., invalid instructions) and (2) whether the inconsistencies detected by code feature disappearance are false positives.

Due to involving extensive manual analysis in this experiment, it is unaffordable to manually check all generated instructions by baselines and all detected inconsistencies by code feature disappearance.
Hence, we used the whole HumanEval dataset and randomly sampled 150 inputs from the APPS test set for manual analysis.
That is, we compared \tech{} and baselines based on the manually-analyzed data.
We also reported the results of \tech{} on the whole test set of APPS based on two exact checking mechanisms (i.e., syntax error and test execution difference) to show its effectiveness more extensively.
Here, we used CodeRL, GPT2 (1.5B), and GPT2 (117M) as the representatives since ChatGPT and Copilot have too high cost on the whole APPS test set.

\subsubsection{Results}
We presented the results from three aspects: the ratio of false positives, the number of detected inconsistencies, and the overlap among the inconsistencies detected by different techniques.

\smallskip
\noindent\textbf{Ratio of false positives:}
Table~\ref{tab:input_fps} presents the percentage of false positives produced by each technique. 
The false positives for baselines are caused by the invalid instructions generated by them.
From Table~\ref{tab:input_fps}, \baselinep{} has the largest percentage of invalid instructions (i.e., 57.32\% on HumanEval and 84.00\% on APPS), which eventually result in false positives.
The main reason lies in that the instructions often contain programming-specific tokens, which are not included in the training data used by \baselinep{}, leading to producing incoherent instructions or missing important information (e.g., identifiers or keywords).
\baselinet{} also has a non-trivial percentage of invalid instructions (i.e., 11.59\% on HumanEval and 13.30\% on APPS), although it can handle programming-specific tokens in input instructions better than \baselinep{} to some extent. 
\tech{} has a small percentage of false positives due to the checking mechanism of code feature disappearance, i.e., only 1.34\% and 0.13\% across all the code generation systems using HumanEval and APPS respectively.
That is, \tech{} significantly outperforms both baselines in producing false positives.

\smallskip
\noindent\textbf{Number of detected inconsistencies:}
Table~\ref{tab:rq1-effectiveness} presents the comparison results among the three techniques in terms of the number of detected inconsistencies after filtering out false positives.
Here, we reported the number of detected inconsistencies by each checking mechanism, and the total number by aggregating all the checking mechanisms.
From Table~\ref{tab:rq1-effectiveness}, \tech{} detects the largest number of inconsistencies among the three techniques on each subject, demonstrating the effectiveness of \tech{} in testing robustness of code generation systems.
Across all subjects, \tech{} shows an average improvement of 466.66\% over \baselinep{} and an average improvement of 104.02\% over \baselinet{}.
Moreover, \tech{} outperforms both baselines under each checking mechanism on almost all the subjects.
Among the three checking mechanisms, test execution difference is the most effective to detect inconsistencies for the three techniques on each subject.
This indicates that test case generation may further facilitate the detection of robustness inconsistencies. 

On the whole APPS test set (including 5,000 inputs), \tech{} detects 1,638, 1,535, and 1,572 robustness inconsistencies based on the two exact checking mechanisms (i.e., syntax error and test execution difference) in CodeRL, GPT-2 (1.5B), and GPT-2 (117M), respectively.
Here, we did not use the checking mechanism of code feature disappearance since it is unaffordable for us to check all inconsistencies detected by it. 
The result further confirms the effectiveness of \tech{} (even with only two checking mechanisms).

\smallskip
\noindent\textbf{Overlap of detected inconsistencies:}
\afterpage{\begin{figure*}[t]
    \center
    \begin{tabular}{c}\hspace{-4.5mm}  \includegraphics[scale=0.35]{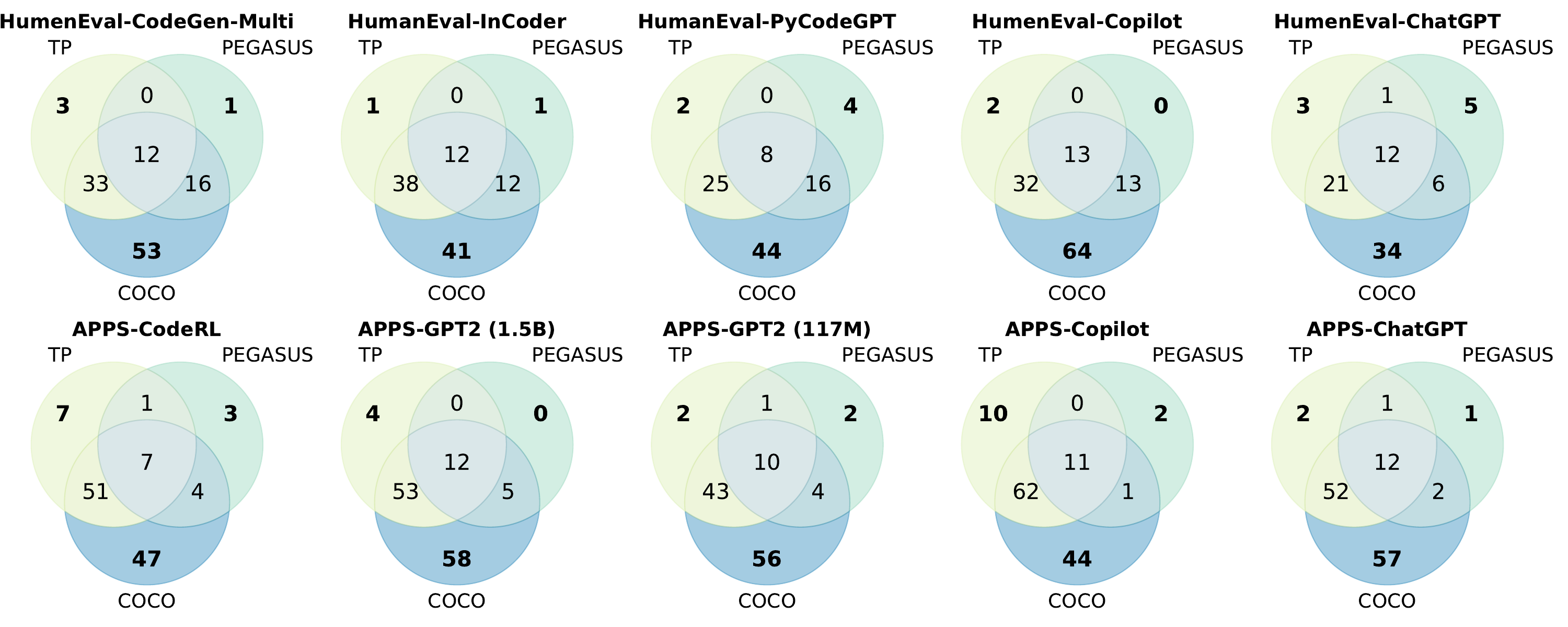}
    \end{tabular}
    \caption{Overlap among the detected inconsistencies by \tech{}, \baselinep{}, and \baselinet{}. 
    }
    \label{fig:overlap}
\end{figure*}}
We further analyzed the overlap of inconsistencies detected by each technique through Venn diagrams shown in Figure~\ref{fig:overlap}.
We found that \tech{} detects the largest number of unique inconsistencies among the three techniques on all the subjects.
On average, 45.43\% of inconsistencies detected by \tech{} across all subjects cannot be detected by both baselines, demonstrating the significantly unique value of \tech{}.
The main reasons for the superiority of \tech{} lie in that both baselines produce too many invalid instructions, and meanwhile \tech{} incorporates a unique checking mechanism (code feature disappearance) to detect inconsistencies.

\finding{\textbf{Answer to RQ1}:
\tech{} is effective in testing robustness of code generation systems, significantly outperforming both \baselinep{} and \baselinet{} by detecting 466.66\% and 104.02\% more semantic inconsistencies, respectively.
Nearly half of (45.43\%) the semantic inconsistencies detected by \tech{} are unique.
In addition, \tech{} does not generate invalid instructions, which is a non-negligible issue for other techniques (i.e., with 11.59\% to 84\% invalid instructions). 
}

\subsection{RQ2: Contribution of Different Code Features and Templates}
\begin{table*}[t]
    \caption{\label{tab:single_muators}Effectiveness of different code features and the corresponding templates. Each cell represents the number of robustness inconsistencies found by aggregating the three checking mechanisms. 
   }
    \begin{adjustbox}{width=1.0\textwidth,center}

\begin{tabular}{l|c|ccccc|ccccc}
\toprule
\multicolumn{1}{c|}{\multirow{2}{*}{\textbf{Level}}} & \multirow{2}{*}{\textbf{Template}} & \multicolumn{5}{c|}{\textbf{HumanEval}} & \multicolumn{5}{c}{\textbf{APPS}} \\ 
\multicolumn{1}{c|}{} &  & \multicolumn{1}{c}{\textbf{CodeGen}} & \multicolumn{1}{c}{\textbf{InCoder}} & \multicolumn{1}{c}{\textbf{PyCodeGPT}} & \multicolumn{1}{c}{\textbf{Copilot}} & \multicolumn{1}{c|}{\textbf{ChatGPT}} & \multicolumn{1}{c}{\textbf{CodeRL}} & \multicolumn{1}{c}{\textbf{GPT2 (1.5B)}} & \multicolumn{1}{c}{\textbf{GPT2 (117M)}} & \multicolumn{1}{c}{\textbf{Copilot}} & \multicolumn{1}{c}{\textbf{ChatGPT}} \\ \midrule
\multirow{2}{*}{\textbf{Dependency}} & \textit{Existence} & 5 & 7 & 5 & 5 & 4 & 14 & 12 & 6 & 22 & 0 \\
 & \textit{Absence} & 60 & 54 & 43 & 77 & 33 & 34 & 49 & 55 & 47 & 68 \\\midrule
\multirow{2}{*}{\textbf{Class}} & \textit{Existence} & 0 & 0 & 0 & 0 & 0 & 1 & 0 & 0 & 0 & 0 \\
 & \textit{Absence} & 63 & 55 & 47 & 85 & 28 & 34 & 58 & 50 & 60 & 66 \\\midrule
\multirow{2}{*}{\textbf{Method}} & \textit{Existence} & 90 & 82 & 78 & 109 & 54 & 20 & 26 & 0 & 30 & 9 \\
 & \textit{Absence} & 0 & 9 & 0 & 10 & 0 & 37 & 53 & 48 & 64 & 66 \\\midrule
\multirow{2}{*}{\textbf{Statement}} & \textit{Existence} & 35 & 55 & 34 & 46 & 30 & 44 & 59 & 52 & 70 & 77 \\
 & \textit{Absence} & 58 & 60 & 54 & 87 & 31 & 35 & 53 & 49 & 70 & 69 \\\midrule
\multirow{2}{*}{\textbf{Expression}} & \textit{Existence} & 64 & 60 & 60 & 73 & 34 & 60 & 67 & 55 & 74 & 87 \\
 & \textit{Absence} & 62 & 59 & 56 & 84 & 37 & 41 & 64 & 65 & 71 & 80 \\\midrule
\multirow{2}{*}{\textbf{Variable}} & \textit{Existence} & 9 & 51 & 14 & 19 & 28 & 65 & 72 & 55 & 67 & 78 \\
 & \textit{Absence} & 53 & 15 & 37 & 56 & 8 & 0 & 0 & 1 & 0 & 0 \\\midrule
\multicolumn{2}{c|}{\textbf{COCO}} & 114 & 103 & 93 & 122 & 73 & 109 & 128 & 113 & 118 & 123 \\ \bottomrule
\end{tabular}
    \end{adjustbox}
    \end{table*}
\subsubsection{Setup}  
To answer RQ2, we analyzed the effectiveness of \tech{} on each subject in terms of the number of detected inconsistencies through each level of code features and each type of templates.
Same as RQ1, we also used the whole HumanEval dataset and the sampled APPS dataset for investigation, and have filtered out the false positives in the inconsistencies detected by code feature disappearance manually.

\subsubsection{Results}
Table~\ref{tab:single_muators} shows the results on \tech{} under each level of code features and each type of templates, where each cell represents the number of detected inconsistencies by aggregating the three checking mechanisms.
From Table~\ref{tab:single_muators}, each level of code features is effective to detect inconsistencies, but detects a smaller number of inconsistencies than that aggregating all the six levels.
This indicates the complementarity of different levels of code features.
Similarly, both existence-based and absence-based templates are also complementary to each other.
We also conducted overlap analysis for them through Venn diagrams, but put them on the project homepage~\cite{coco_homepage} due to limited space.

We found that statement-level and expression-level code features are more effective to detect inconsistencies than the other levels, which contribute to 64.74\% and 74.70\% of the detected inconsistencies on average across all subjects.
This is because the two levels of code features are more common in source code (also in generated code), which can provide \tech{} more opportunities to generate effective concretized instructions to detect inconsistencies.

In addition, absence-based templates are much more effective to detect inconsistencies than existence-based templates for dependency-level and class-level code features.
This is because it is rare for HumanEval and APPS datasets to generate code with class and dependency information, leading to the inapplicability of existence-based templates in most cases.
Another observation from Table~\ref{tab:single_muators} is that method-level code features perform very differently on the subjects using HumanEval and those using APPS.
Specifically, method-level code features work much better when using existence-based templates for the subjects using HumanEval, while those work much better when using absence-based templates for the subjects using APPS.
The reason is similar, i.e., it is rare for APPS to generate code with method definitions while it is relatively common for HumanEval.
\finding{\textbf{Answer to RQ2}: 
Different levels of code features and different types of templates are complementary to detect inconsistencies for code generation systems.
Expression-level code features are the most effective, contributing up to 74.70\% of the detected inconsistencies on average across all subjects.
}

\subsection{RQ3: Robustness Enhancement}

\subsubsection{Setup} 
In RQ3, we investigated the value of concretized instructions generated by \tech{} by using them to enhance robustness of code generation systems.
Following the existing work~\cite{Chen2021QAASKER,shen2022qaqa}, we adopted the widely-used fine-tuning strategy to enhance robustness.
However, HumanEval is a benchmark containing only 164 artificially-constructed problems and solutions. Such a small number of samples may be insufficient to fine-tune LLMs.
Hence, we investigated RQ3 on only the APPS dataset, which has been officially divided into training data and test data. 
Considering model availability, we used GPT2 (117M), GPT2 (1.5B), and CodeRL as the representatives in this experiment.
Here, we denoted the training set of APPS dataset as $S_1$ and the test set as $S_2$.
$S_1$ is used to construct the corresponding fine-tuning set, while $S_2$ is regarded as the corresponding evaluation set for the fine-tuned code generation system to avoid data leakage.

For each input in each $S_1$, we applied \tech{} to randomly construct three concretized instructions, which pair with the original instruction as fine-tuning data.
That is, it is used to fine-tune the target code generation system.
To ensure the quality of labels in the fine-tuning set, \tech{} extracts code features from the ground-truth code corresponding to an input instruction to generate a concretized instruction, and assembled the concretized instruction with the ground-truth code as a fine-tuning instance.
Subsequently, we applied \tech{} to evaluate the robustness of each fine-tuned code generation system based on the corresponding $S_2$.
We collected the testing results to check whether the robustness of code generation systems is improved with the aid of \tech{}.

% \begin{table}[t]
%     \caption{\label{tab:retrain}Performance of the original model and the fine-tuned model, as well as the number of inconsistencies detected by \tech{}.}
%     % \vspace{-1mm}
%     \begin{adjustbox}{width=0.5\textwidth,center}
%     \begin{threeparttable}

%     \begin{tabular}{l|cc|cc}
%     \toprule
%     \multirow{2}{*}{\textbf{Type}} & \multicolumn{2}{c|}{\textbf{HumanEval-CodeGen}} & \multicolumn{2}{c}{\textbf{APPS-GPT2 (117M)}} \\
    
%      & \textbf{CodeBLEU} & \textbf{\# Incons.} & \textbf{CodeBLEU} & \textbf{\# Incons.} \\\midrule
%     \textbf{Original} & 0.202 & 48 & 0.288 & 113 \\
%     \textbf{Fine-tuned} & 0.208 & 12 & 0.287 & 82 \\ \bottomrule
% \end{tabular}

%     \end{threeparttable}
%     \end{adjustbox}

% \end{table}

\begin{table}[t]
    \caption{\label{tab:retrain}Performance of the original model and the fine-tuned model, as well as the number of inconsistencies detected by \tech{}.}
    % \vspace{-1mm}
    \begin{adjustbox}{width=0.5\textwidth,center}
    \begin{threeparttable}

    \begin{tabular}{l|cc|cc}
    \toprule
    \multirow{2}{*}{\textbf{Model}} & \multicolumn{2}{c|}{\textbf{CodeBLEU}} & \multicolumn{2}{c}{\textbf{\#Inconsistencies}} \\ 
     & \textbf{Original} & \textbf{Fine-tuned} & \textbf{Original} & \textbf{Fine-tuned} \\\midrule
    \textbf{GPT2 (117M)} & 0.288 & 0.287 & 113 & 82 \\
    \textbf{GPT2 (1.5B)} & 0.255 & 0.249 & 128 & 59 \\
    \textbf{CodeRL} & 0.268 & 0.252 & 109 & 89 \\ \bottomrule
    \end{tabular}
    
    \end{threeparttable}
    \end{adjustbox}

\end{table}

\subsubsection{Results} 

Table~\ref{tab:retrain} shows the number of detected inconsistencies on the original code generation system and the fine-tuned code generation system, respectively.
Here, we reported the number of inconsistencies by aggregating all the three checking mechanisms after filtering out false positives caused by code feature disappearance (Column ``\# Inconsistencies'').
For all the subjects, \tech{} indeed enhances robustness.
Specifically, the number of inconsistencies detected on the original GPT2-1.5B is 128, while that on the fine-tuned GPT2-1.5B is only 59.
Similarly, the number of inconsistencies detected on the original CodeRL is 109, while that on the fine-tuned CodeRL is 89.
The improvement brought by \tech{} in terms of the number of inconsistencies ranges from 18.35\% to 53.91\% across all three subjects.
Following the existing work~\cite{Wang2021codet5}, we also measured the performance of code generation systems before and after fine-tuning by calculating the average CodeBLEU score between generated code and ground truth (Column ``CodeBLEU'').
We found that the fine-tuning strategy with concretized instructions generated by \tech{} does not affect code-generation performance obviously, while largely enhancing robustness.

\finding{\textbf{Answer to RQ3}: 
The concretized instructions generated by \tech{} can help reduce 18.35\% to 53.91\% robustness inconsistencies with fine-tuning.}

\section{Threats to Validity}\label{sec:threats}
The threats to \textit{internal} validity mainly lie in the implementation of \tech{} and manual analysis. 
To reduce the first threat, two authors checked all our code and wrote test cases at the programming stage. 
To reduce the second threat, two authors independently validated each input generated by baselines and each inconsistency detected by code feature disappearance.
Once their validation results are different, the third author discussed with them to determine the final result.

The threat to \textit{external} validity mainly lies in the subjects used in our study. 
To reduce this threat, we carefully considered the diversity of both code generation systems and datasets.
Moreover, these subjects have been widely used in the existing work~\cite{Mastropaolo2023copilot,li2022cctest}.
In the future, we will extend \tech{} to more programming languages and evaluate it on more subjects. 
Additionally, the number of test cases in the dataset may affect the checking of semantic consistency, i.e., a small number of test cases may not reveal the test execution differences between the original and generated code. We have used the augmented test set of HumanEval in this study to minimize the threat. In the future, we will conduct test case augmentation on APPS.

The threats to \textit{construct} validity mainly lie in randomness and parameter settings. 
To reduce the first threat, we investigated the influence of randomness on \tech{} by repeating the experiment five times on six subjects (excluding Copilot and ChatGPT due to huge costs).
The remaining experimental setup was consistent with RQ1. 
We calculated the standard deviation of the number of detected inconsistencies across the five repeated experiments, which just ranges from 0.80 to 2.89 across the six subjects.
The results show the little influence of randomness on \tech{}, demonstrating its stable effectiveness.
To reduce the second threat, we studied the influences of main parameters on \tech{}, which will be presented in Section~\ref{subsec:parameters}.

\section{Discussion}\label{sec:discussion}

\subsection{Efficiency of \tech{}}\label{subsec:effectiveness}

We analyzed the efficiency of \tech{} in generating new instructions for testing code generation systems by taking ChatGPT with HumanEval as the representative.
On average, \tech{} took 1.592 milliseconds to construct a concretized instruction, among which 0.7 millisecond was spent on code feature extraction and  0.892 millisecond was spent on concretized instruction construction.
In particular, \tech{} does not rely on other heavy tools (such as a sequence-to-sequence model for \baselinep{} and a translation engine for \baselinet{}) to generate new instructions, and thus consumes less machine resource.
Also, \tech{} does not generate invalid instructions that are non-neglectable for baselines.
Overall, \tech{} is able to efficiently test code generation systems.

\subsection{Influences of Hyper-parameters}\label{subsec:parameters}

\begin{figure}[t]
    \centering
    \subfloat[$m$]{
        \begin{minipage}[b]{0.23\textwidth}
            \includegraphics[width=1\textwidth]{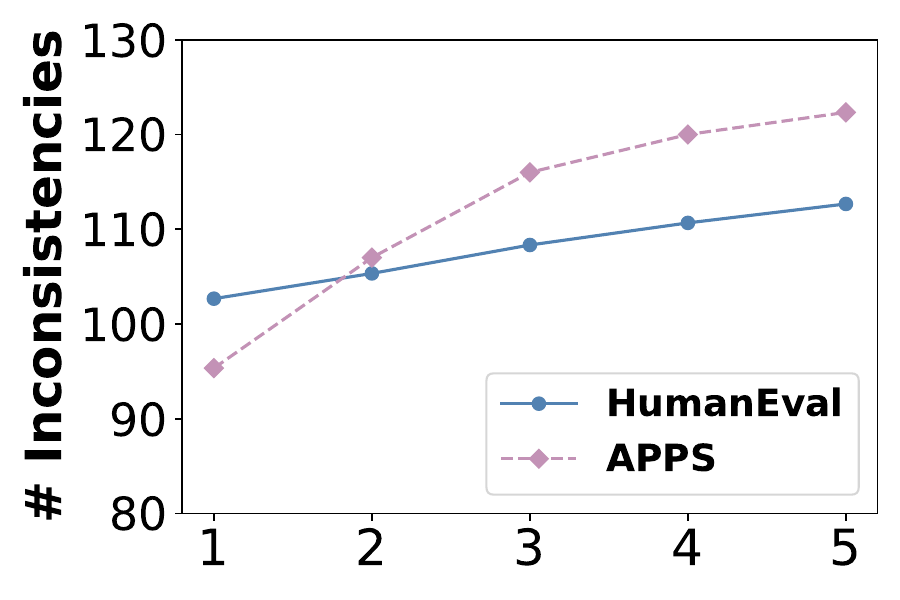}
        \end{minipage}
    }
    \hspace{-2mm}
    \subfloat[$n$]{
        \begin{minipage}[b]{0.23\textwidth}
            \includegraphics[width=1\textwidth]{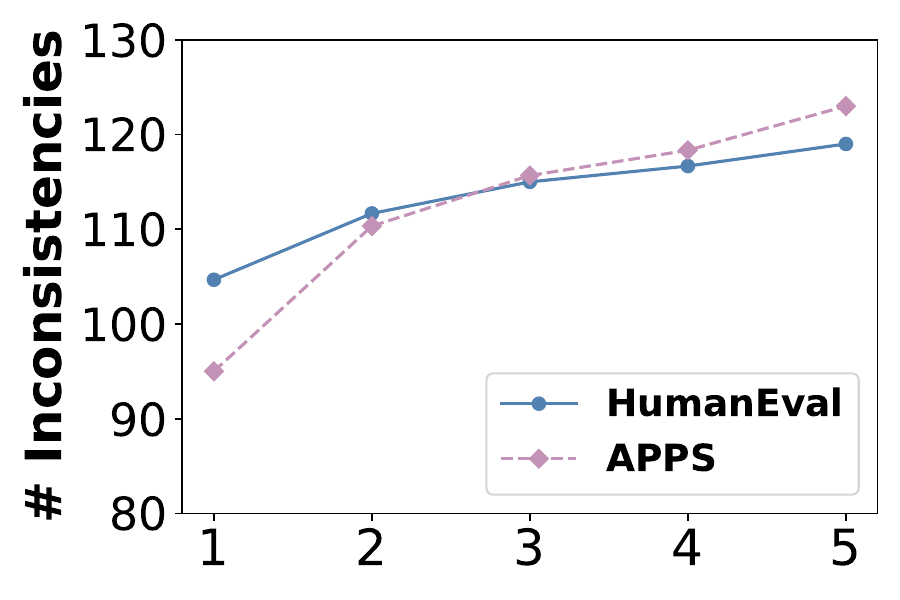}
        \end{minipage}
    }
    \caption{Influences of $m$ and $n$ on the effectiveness of \tech{}.}.
    \label{fig:parameters}
\end{figure}

As introduced in Section~\ref{subsec:implementations}, we studied the influences of two hyper-parameters (i.e., $m=\{1,2,3,4,5\}$ and $n=\{1,2,3,4,5\}$).
When changing a hyper-parameter, another hyper-parameter is set to the default setting.
Here, $n$-order concretization refers to performing first-order concretization on each instruction generated via $(n-1)$-order concretization.
Figure~\ref{fig:parameters} shows the effectiveness of \tech{} under different settings of $m$ and $n$.
Due to limited space, we reported the total number of detected inconsistencies across all the code generation systems using the HumanEval dataset and the APPS dataset, respectively.
With the settings of $m$ and $n$ increasing, the effectiveness of \tech{} can be improved. 
This is as expected, since the input space of \tech{} is large, exploring it more sufficiently can have more opportunities to generate inconsistency-triggering instructions.
However, this process can incur more costs.
The results indicate that more efficiently exploring the input space is a promising direction to further improve \tech{}.
Since our contribution is to design the idea of concretizing instructions for testing robustness of code generation systems, rather than better explore the input space, we set $m=1$ and $n=1$ by default in \tech{} currently.

\subsection{Comparison in Terms of Code Similarity}\label{subsec:code_bleu}

\begin{figure}[t]
    \center
    \begin{tabular}{c}\hspace{-4.5mm}  \includegraphics[scale=0.35]{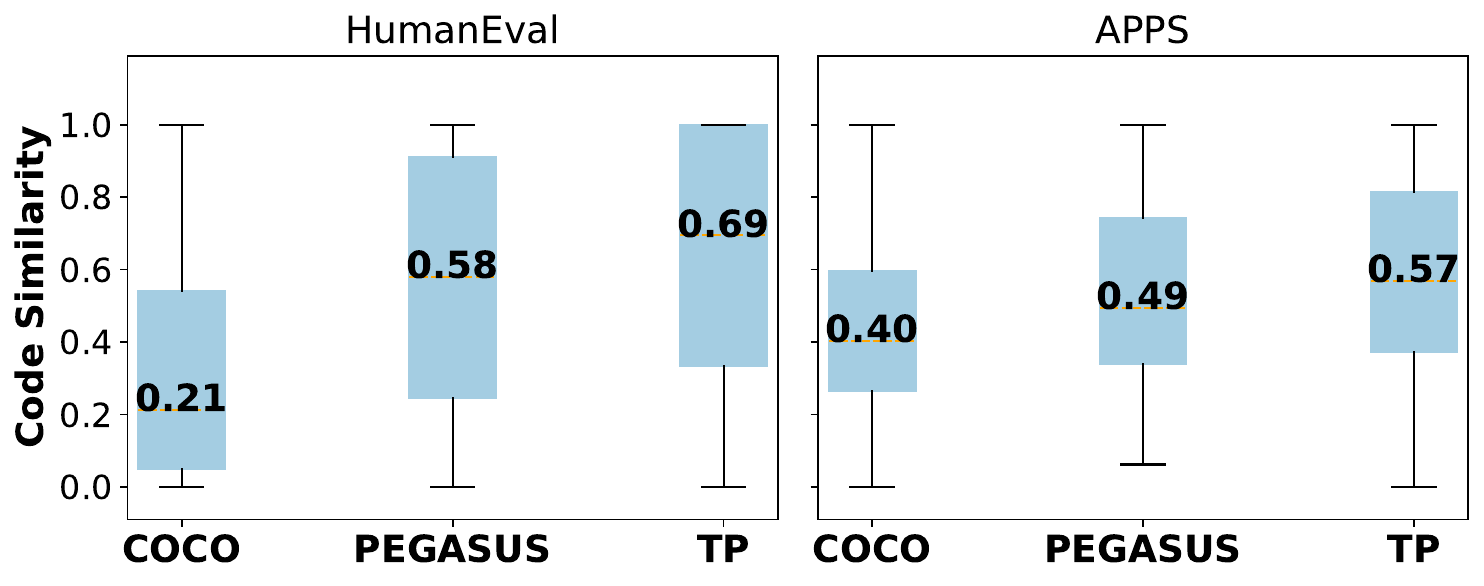}
    \end{tabular}
    \caption{Code similarity comparison in terms of CodeBLEU among \tech{}, PEGASUS, and \baselinet{}.}
    \label{fig:code_bleu}
\end{figure}

The existing work measured code similarity in terms of CodeBLEU~\cite{Ren2020CodeBLEU} to test robustness of code generation systems~\cite{Mastropaolo2023copilot}.
Smaller CodeBLEU values (i.e., smaller code similarity) mean that it is more likely to detect semantic inconsistencies.
Since this metric cannot exactly determine the number of detected inconsistencies, we did not include it in our main experiments.
As complementary, we also used it to compare \tech{} and the baselines here.

Figure~\ref{fig:code_bleu} shows the comparison results among the three techniques in terms of code similarity between the generated code corresponding to the pair of an original instruction and the generated instruction.
It uses box-plots to show the distribution of CodeBLEU scores across all the input pairs constructed by each technique across all the subjects using HumanEval and APPS, respectively.
We found that the median CodeBLEU score of \tech{} is smaller than those of both baselines in each figure, further confirming the effectiveness of \tech{}.
On HumanEval, the median CodeBLEU score of \tech{} across all subjects is 0.21, while those of \baselinep{} and \baselinet{} are 0.57 and 0.69, respectively.
That is, \tech{} is more effective to mislead the code generation system to produce more different code from the original one.

\section{RELATED WORK}

The most related work to ours is the study performed by Mastropaolo et al.~\cite{Mastropaolo2023copilot}, which is the first to evaluate robustness of code generation systems (i.e., Copilot).
Specifically, they investigated whether the testing techniques for general text-to-text software (\baselinep{} and \baselinet{}) can detect robustness inconsistencies in code generation systems.
Both of them paraphrase a given instruction to a semantic-preserving one for testing.
Different from it, our work proposes a novel testing technique \tech{} for code generation systems.
% \tech{} does not paraphrase the original instruction, but complements more details to it by extracting code features in the original code.
Instead of paraphrasing the original instruction, \tech{}  complements more details to it by extracting code features in the original code.
Our study has demonstrated the superiority of \tech{} over both \baselinep{} and \baselinet{}.
% In particular, our study has shown its superiority over \baselinep{} and \baselinet{}.

Besides, there are several works focused on testing other kinds of code models~\cite{Pour2021search,Zhou2022codecomments,li2022cctest,zhang2020MHM,yang2022alert}.
For example, Li et al.~\cite{li2022cctest} proposed CCTEST to test \textit{code completion systems} by defining a set of equivalent code transformation rules.
Zhang et al.~\cite{zhang2020MHM} and Yang et al.~\cite{yang2022alert} proposed to test \textit{code classification models} by renaming identifiers.
All of them target the systems taking source code as input and design code transformation rules to generate new inputs, while our work targets code generation systems, whose inputs are natural language instructions.
Indeed, \tech{} designs the concretization process to generate new instructions for testing.

% focus on code completion or code representation systems, where the input is source.

\section{Conclusion}
In this work, we propose a novel technique \tech{} to test the robustness of code generation systems by exploiting the characteristics of this kind of systems.
Specifically, it makes the original instruction more concrete by incorporating features known to be contained in the original code.
For a robust code generation system, concretizing instructions with known outputs should not change the original code semantics.
The concretization process conforms to the usage scenario of code generation systems in practice by providing more detailed programming requirements.
By conducting a study on eight code generation systems, \tech{} is demonstrated effective to detect robust inconsistencies, outperforming baselines significantly.
In particular, the robustness of code generation systems can be largely enhanced by fine-tuning with the concretized instructions by \tech{}.
Note that our work does not aim to detect all robustness issues in code generation systems with \tech{}.
Indeed, \tech{} depends on the features in generated code, which may limit the diversity of features.
In the future, we can incorporate more information to enhance the robustness testing of code generation systems.

% \newpage

\balance
\bibliographystyle{ACM-Reference-Format}
\bibliography{ref.bib}

\end{document}